# A Personalised User Authentication System based on EEG Signals.

Christos Stergiadis[1,2], Vasiliki-Despoina Kostaridou[1], Simeon Veloudis[3], Dimitrios Kazis[4], and Manousos Klados[1,2]

[1]Department of Psychology, University of York Europe Campus, City College, Thessaloniki, Greece

[2]Neuroscience Research Center (NEUREC), University of York Europe Campus, City College, Thessaloniki, Greece

[3]Department of Computer Science, University of York Europe Campus, City College, Thessaloniki, Greece

[4]3rd Department of Neurology, Aristotle University of Thessaloniki, Exochi, 57010 Thessaloniki, Greece

## Abstract

Conventional biometrics have been employed in high security user authentication systems for over 20 years now. However, some of these modalities face low security issues in common practice. Brain wave based user authentication has emerged as a promising alternative method, as it overcomes some of these drawbacks and allows for continuous user authentication. In the present study we address the problem of individual user variability, by proposing a data-driven Electroencephalography (EEG) based authentication method. We introduce machine learning techniques, in order to reveal the optimal classification algorithm that best fits the data of each individual user, in a fast and efficient manner. A set of 15 power spectral features (delta, theta, lower alpha, higher alpha, and alpha) is extracted from the three EEG channels. The results show that our approach can reliably grant or deny access to the user (mean accuracy 95,6%), while at the same time poses as a viable option for real time applications, as the total time of the training procedure was kept under one minute.

## 1. Introduction

As modern society has already transited into the era of information and technology, security and privacy are becoming increasingly important. The need for effective user authentication systems in a trusted and autonomous manner is more evident than ever, in order to prevent intruder attacks or information leaks. The potential of using Electroencephalography (EEG) signals as a biometric tool for person authentication is an area that attracted increased attention over the last decade [1]–[12], as it overcomes some of the deficiencies that the already established methods do present, like ensuring the liveliness of the user and allowing for continuous user authentication [13] .

Biometrics in general are defined as the unique behavioral or physiological characteristics that can be used for the identification of a person [14]. Authentication systems using well-established biometric signals like fingerprint, iris, voice, and gait recognition regularly face low-security issues, as they are vulnerable to spoofing tools [5]. Such tools can be the artificially generated "gummy" fingers [15] for fooling fingerprint recognition, voice coders for voice recognition, contact lenses for iris recognition, and adversarial attacks for gait recognition [16]. In addition to these issues, conventional biometric-based authentication systems may give rise to violent attacks in which the attacker forces the victim to provide his/her biometric traits to the system (e.g. at gunpoint threat, using a dismembered finger, etc). Other means of authentication, like password-based techniques (something that the person knows), may be easy to use but are also threatened by malicious attacks like the popular dictionary or brute force attack, and the exploitation of user mistakes [13], while also being vulnerable to social engineering related attacks. Additionally, token-based authentication, which is connected with something that the user possesses like for example a key, a card, or a USB Dongle [17] can be proven inconvenient as the user needs to carry the token every time that he/she requires access. Moreover, there is also the danger of the object being stolen or mimicked by reverse engineering techniques [17].

A user authentication method based on brain wave activity can address the aforementioned drawbacks or complement them and also provide solutions to high-level security systems and continuous authentication requirements [18]. Electroencephalography (EEG) is a highly individualistic biometric that has high inter-subject variability and low intra-subject variability [12]. Therefore, it can be used for the efficient identification and authentication of a user and ensure shielding against intruders. Moreover, brain waves are an intrinsic characteristic that makes EEG strong against mimicking and identity theft, in contrast with conventional biometrics stemming from the human body. In fact, an attacker cannot force the user to authenticate himself, as stress and pressure seriously affect the EEG signals [5], while also the liveliness of the user is ensured. Another important key factor of EEG is that it features cognitive processes that can be detected unconsciously, providing the opportunity for continuous user authentication.

During the last decade, a large number of publications have emerged, dealing with EEG-based authentication techniques [1]–[12]. These works naturally strive to optimize the accuracy and ease of use of the proposed approaches by typically relying on an efficient combination of the EEG features that are chosen to represent the individual's brain activity on one hand, and the classifier that is used for the classification and the final decision of the system to grant or deny access on the other. Pham et al. [13] used Autoregressive (AR) linear parameters and Power Spectral Density (PSD) components (3-80Hz) which were fed into a Support Vector Machine (SVM) classifier, achieving Equal Error Rate (EER) as low as 0.002. The same authors later investigated the same combination in a different frequency band (1-30Hz) achieving an authentication system with 97% accuracy [19].

Both AR and PSD are widely used in EEG authentication studies. Poulos et al. [20] combined AR features with Kohonen's Vector Quantizer (VQ) as the classifier, while Paranjape et al. [21] classified the same type of features with discriminant analysis algorithms. In more recent studies, Thomas et al. [22] experimented with resting-state Eyes Open (EO) and Eyes Closed (EC) EEG, extracting PSD features in the gamma band (30-50 Hz) and used specific thresholds in order to grant (or deny) access to the user, finally achieving EER=0.019. Additionally, in many studies, external stimuli or specific imaginary tasks were presented to the user and the utility of specific features of the Event Related Potentials (ERPs) was assessed. For example, Valsaraj et al. [14] analyzed the EEG data, looking for characteristic features in the ERPs that were elicited from Motor Imagery (MI) and real movements. By combining different MI actions and using AR and PSD, their proposed authentication method reached an accuracy of 98.28%.

Another direction of research in this field uses the Neural Network (NN) classifier for human identification and classification. From very early-stage studies, Poulos et al. [23] employed the learning vector quantizer (LVQ) and spectral features reaching accuracies ranging from 80% to 100%. Subsequently, the same researchers experimented on the same classifier but with AR and bilinear model features [24], resulting in accuracies from 56% to 80%.

Later on, the back-propagation and the feed-forward Neural Networks gained prominence, with Hema and Osman [25] achieving average accuracy between 80% and 90% by using PSD features and feed-forward NN for their classification. Mu and Hu [26] reached 80% in accuracy in their authentication system by choosing AR and Fischer distance as the features and a back-propagation (BP) NN classifier. Fischer distance has also been drafted, in later studies, along with fuzzy entropy as the representative features when visual stimuli (self-photos vs non-self-photos) were introduced to the subject [5], and with the use of BP Neural Network as the classification theme, accuracies of 87.30% were achieved, with False Acceptance Rate (FAR) at 5.50% and False Rejection Rate (FRR) at 5,60%.

Most of the current works have chosen a specific combination of targeted features and a classifier in order to complete their analysis and achieve optimal accuracy. Nevertheless, none of the proposed methods is neither user specific, nor it takes into consideration the specific, each time, characteristics of the given dataset. This may reduce the performance of the implemented analysis and affect the classification accuracy. Hence, a practical method that addresses this problem needs to be devised if we consider EEG signals as a viable option for real-time human authentication, with advanced accuracy and reliability. The present study proposes an EEG-based authentication system built around the spectral features that characterize brain activity, and the implementation of a Machine Learning (ML) algorithm for the classification of these features. The Auto-WEKA software [27] is used in order to ensure that the used algorithms are the most suitable for our data. More specifically, 15 power spectral density features are extracted from 3 central electrodes (Fz, Cz, Pz) in five different frequency bands, for 15 subjects. The choice for the best descriptive features as as well as the optimal ML algorithm based on the aforementioned features is then automatically appointed by running Auto-WEKA.

The rest of the paper is organized as follows. Section 2 presents the dataset, the equipment the features and the classification methods used in the study. In Section 3, we present the results for the 15 subjects tested herein, while in Section 4 we provide a discussion of our findings in light of the current literature and a conclusion of our work.

## 2. Dataset and experimental setup

### 2.1 Participants

The sample of participants consisted of 15 individuals, and more specifically, 8 males and 7 females with a mean age of 23.2 ± 5.5 for the males and 21.2 ± 3.4 for the females. The criteria that were set for exclusion from participation concerned the history of neurological and psychiatric illness, substance abuse history, medication, and any other coexisting factors that could affect the brain's neurophysiology. All participants had normal or corrected to normal vision and were asked to not consume alcohol or caffeine the day prior to their participation. All the experiments were performed at the same time, to the best possible extent. This research was approved by the Ethical Committee of Aston University. The participation was anonymous and confidential.

## 2.2 EEG signal acquisition

EEG measurements were recorded from 128 electrodes distributed across the scalp according to the EGI Geodesic EEG System (GES) and with reference electrodes positioned at the mastoids. The correct EEG Geodesic net size was soaked in a saline solution for 5 minutes and applied to their head. The recordings were performed with an EGI GES 300 system. All electrode impedances were maintained at less than 5 k$\Omega$. The sampling rate for all measurements was 250 Hz (using Net Station 4.3 software on an Apple Macintosh Dual 2 GHz Power PC G5, Mac IOS X Version 10.4.4). After each experiment, the nets were disinfected in order to maintain high hygienic standards. The data used for the purposes of this work were recorded during resting state with eyes opened (30 seconds where participants were asked to keep their eyes open and fixate on a cross appearing at the center of the screen) in order to simulate a real scenario of user identification and only Fz, Cz and Pz were further used.

## 2.3 Preprocessing and Feature Extraction

All EEG signals were referenced according to the linked earlobe montage [28], filtered at the frequency range of 0.5 – 40 Hz, and submitted to the Adaptive Mixture ICA (AMICA) algorithm [29], and the REG ICA [31], [32] methodology was used to clean the independent components from artifacts. Specifically ocular artifacts, which are the most troublesome in eyes open conditions, were cross-validated by using the using EEGLAB's [33] toolbox called ICLABEL [34].

The cleaned EEG signals were divided into 500 random segments of 4 seconds each, in order to increase the external validity of the current study. For each segment, the power of five different brain waves was computed (delta [0-4Hz], theta [4-8Hz], lower alpha [8-10Hz], higher alpha [10-12Hz], and alpha [8-12Hz]). This resulted in a feature-set of 15 (3 channels x 5 bands) features per subject with 500 instances.

## 2.4 Classification procedure

In order to replicate a realistic scenario, classification was performed on an individual basis. Thus, for each participant (user), a separate dataset of 1000 instances were formed, where half of the instances were derived from the 500 random segments from the user, and the other 500 were randomly chosen from the remaining 14 participants. In an ideal scenario, a robust user authentication algorithm should grant access to the 500 instances that come from the user and deny access to the rest 500 instances coming from the other 14 users. This procedure resulted in 15 different user-specific datasets that were further imported into WEKA [35], where Auto-WEKA [27] was used with 1 minute as the time limit for training. The default time limit was reduced from 15 mins to only 1 min having in mind the usability of our approach (**Figure 1**).

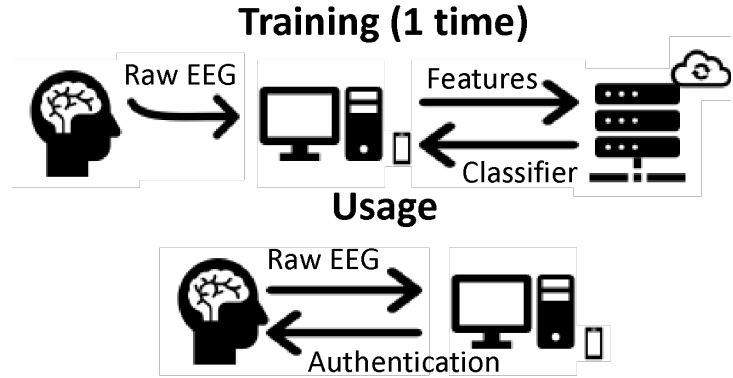

**Figure 1** *System Illustration of the two different phases of the classifier. First, there is a training phase, where the server computes the optimal classification algorithm, and then the system grants or denies access to the user based on this classifier.*

In the proposed system, the user wears a wireless and wearable EEG headset that sends the raw EEG signals (~30 seconds) to the terminal authentication app. The app computes the 15 features with 500 instances mentioned above which are uploaded to the server. The server forms a dataset consisting of the 500 instances of the user and a random selection of other 500 instances from other users that are stored in the database. Then, the server computes the optimal, user-specific, classification algorithm and returns it to the terminal app. The aforementioned procedure lasts around 1 minute and it runs only once (upper schema). Since the classifier is saved in the terminal app, the app can grant or deny access to the user using the provided EEG data.

## 3. Results

Table 1 presents the authentication results after the proposed method was implemented on the 15 subjects. Mean accuracy of 95,6% was obtained for all subjects, while the sensitivity of the system was 0,93 (±0,04) and the specificity was 0,98 (±0,02). The fluctuation of the latter two values across the subjects is depicted in Figure 2.

| SUBJECT | GRANT ACCESS | | DENY ACCESS | | Sensitivity | | Specificity | | Accuracy | Kappa |
|---|---|---|---|---|---|---|---|---|---|---|
| | GRANTED | DENIED | GRANTED | DENIED | TPR | FPR | TNR | FNR | | |
| SS01 | 500 | 0 | 0 | 500 | 1 | 0 | 1 | 0 | 100% | 1 |
| SS02 | 436 | 64 | 16 | 484 | 0.872 | 0.032 | 0.968 | 0.128 | 92% | 0.84 |
| SS03 | 493 | 7 | 25 | 475 | 0.986 | 0.05 | 0.95 | 0.014 | 97% | 0.936 |
| SS04 | 420 | 80 | 49 | 451 | 0.84 | 0.098 | 0.902 | 0.16 | 87% | 0.742 |
| SS05 | 474 | 26 | 7 | 493 | 0.948 | 0.014 | 0.986 | 0.052 | 97% | 0.934 |
| SS06 | 450 | 50 | 8 | 492 | 0.9 | 0.016 | 0.984 | 0.1 | 94% | 0.884 |
| SS07 | 475 | 25 | 9 | 491 | 0.95 | 0.018 | 0.982 | 0.05 | 97% | 0.932 |
| SS08 | 476 | 24 | 14 | 486 | 0.952 | 0.028 | 0.972 | 0.048 | 96% | 0.924 |
| SS09 | 469 | 31 | 8 | 492 | 0.938 | 0.016 | 0.984 | 0.062 | 96% | 0.922 |
| SS10 | 482 | 18 | 7 | 493 | 0.964 | 0.014 | 0.986 | 0.036 | 98% | 0.95 |
| SS11 | 466 | 34 | 9 | 491 | 0.932 | 0.018 | 0.982 | 0.068 | 96% | 0.914 |
| SS12 | 472 | 28 | 6 | 494 | 0.944 | 0.012 | 0.988 | 0.056 | 97% | 0.932 |
| SS13 | 465 | 35 | 8 | 492 | 0.93 | 0.016 | 0.984 | 0.07 | 96% | 0.914 |
| SS14 | 478 | 22 | 0 | 500 | 0.956 | 0 | 1 | 0.044 | 98% | 0.956 |
| SS15 | 451 | 49 | 8 | 492 | 0.902 | 0.016 | 0.984 | 0.098 | 94% | 0.886 |

*Table 1* Authentication results. The classification accuracy, sensitivity, specificity, and Kappa index across the 15 subjects for the 500 instances coming from the EEG of each user and the 500 randomly chosen instances coming from the remaining 14 subjects, which should be treated as the impostor signals.

The highest accuracy was obtained for the first subject (100%), while the lowest levels of accuracy were observed for the fourth subject (87%), which had a sensitivity of 0.84 and specificity of 0.902.

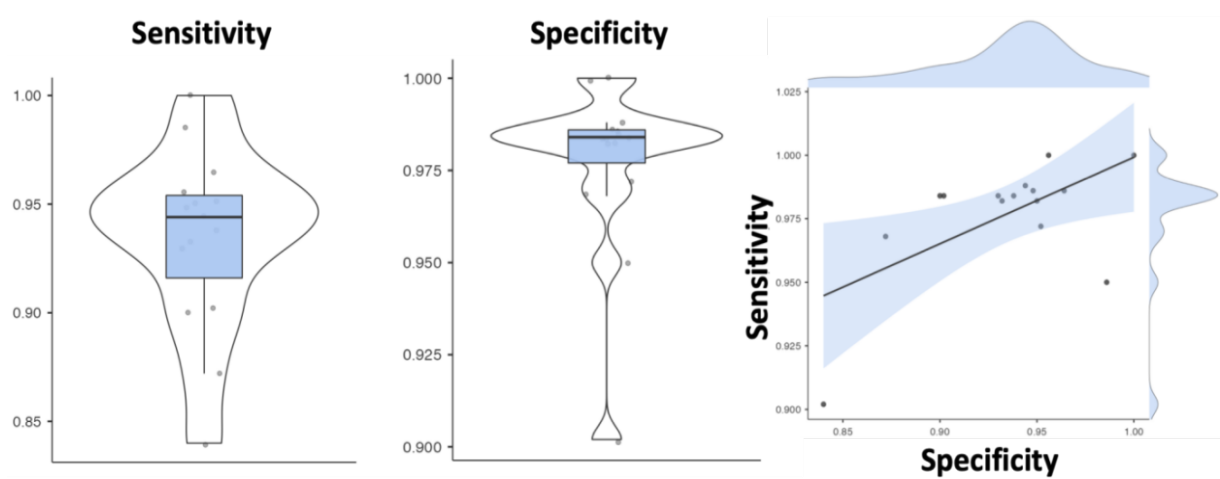

*Figure 2* Classification performance. The violin and the box plots describe the distribution of sensitivity and specificity across the 15 subjects, while from the scatterplot, it is obvious that one subject was an outlier indicating that our results would be much better by excluding this subject.

Furthermore, we employed statistical analysis, although the very obvious results, in order to prove our approach statistically. The accuracy, FPR and FNR were tested using one sample t-test or one sample Wilcoxon W test if the variable is not normally distributed. According to Saphiro-Wilk test of normality only the FPR is normally distributed (W=0.944, p=0.437), while accuracy (W=0.835, p=0.011) and FNR (W=0.705, p=0.001) are not normally distributed. So for the FPR, one sample t-test revealed that our approach has significantly lower FPR (M =0.06, SD = 0.04) than random guessing, t(14) = -40.4, p =0.001. On the other hand, the W test for accuracy revealed that our approach has significantly higher accuracy (M=95.6%, SD=2.9%) than the random guessing W(14)=120, p=0.001, while the FNR was significantly lower (M =0.02, SD = 0.01) than the random guessing W(14)=0, p=0.001.

## 4. Discussion and Conclusion

The usage of EEG signals in human authentication systems has proven to be a very effective technique as it overcomes most of the drawbacks that conventional biometric tools, like iris, fingerprint, or voice-based applications can present in everyday practice. Some of the advantages of using brain waves include their suitability for continuous user authentication systems, as they pose as nonconscious biometrics, and the fact that they can be utilized in high-level security facilities, ensuring the liveliness of the person requiring access and averting intruder attacks.

| Paper | No. of subjects | No. of EEG channels | Features | Accuracy |
|---|---|---|---|---|
| [19] | 40 | 8 | AR linear parameters and PSD components (1-30Hz) | 97,10% |
| [36] | 10 | 2 (Fp1 &Fp2) | Fuzzy entropy and Fisher distance | 87,30% |
| [12] | 8 | 9 | Low-frequency SSVEP components | 96.78% |
| [37] | 10 | 10 | Wavelet Packet Decomposition and Correlation-based features | 95% |
| [3] | 32 | 1 | Wavelet based (time-frequency) features: 1)mean 2)standard deviation 3)entropy for the wavelets of the five frequency bands | 94,04% |
| [14] | 25 | 4 | AR linear parameters, PSD components | 98,28% |

| Ref | | | Features | Accuracy |
|---|---|---|---|---|
| [1] | 5 | 14 | 1) AR coefficients 2) PSD components 3) total power 4) interhemispheric power differences 5) interhemispheric linear complexity | 97,69% |
| [11] | 10 | 18 + 5 subject-specific channels | The difference between the averaged signals in response to self-face | 86,10% |

*Table 2 Previous EEG based user authentication studies.*

Despite the increasing number of studies in this field and the numerous different approaches (Table 2), from trying to find the optimal combination of features (AR models, PSD components, etc.) and classification algorithms (LDA, SVM, CNN, etc.), to the use of innovative measures like fuzzy entropy and exploiting eye blinking signals, no studies to the best of our knowledge have taken into account the individual characteristics of the dataset in use. In this paper, we computed the PSD of the EEG signal in five different bands (delta [0-4Hz], theta [4-8Hz], lower alpha [8-10Hz], higher alpha [10-12Hz], and alpha [8-12Hz]) from 3 central electrodes (Fz, Pz, and Cz) resulting in 15 different features for each subject.

The novelty of our method is that we introduced the use of the machine learning algorithm "WEKA" as an extra step after the feature extraction stage, in order to appoint the optimal classification algorithm for each individual. As so, we achieved an overall mean accuracy of 95,6%, a sensitivity of 0,93 and a specificity of 0,98 across the 15 subjects. Another important feature of our proposed methodology is that the EEG signal was recorded from only 3 central electrodes and for just 30s, making the system very efficient. In addition, the total time for the EEG signal recording, the feature extraction, and the auto-WEKA based algorithm selection for the classification procedure was kept under one minute, as this study reflects the limitations and implications of real life practices and aims to provide solutions for practical use of such EEG based authentication systems.

Finally, regarding future directions, EEG-based user authentication has some general issues that need to be addressed. For example, if the user is not interested in getting authenticated then his/hers brain waves can be altered leading to failures of the authentication system. Also, regarding our approach and proposed methodology, further datasets need to be examined, with a larger number of participants, and the possible use of some of the recently emerged and increasingly used deep learning algorithms can also provide higher accuracies despite the probably increased computational time cost.